\documentclass[aps,prd,superscriptaddress,amsmath,amsfonts,groupedaddress,10pt,english]{revtex4}
\usepackage{amssymb,amsmath,epsfig}
\usepackage{graphicx}
\usepackage{float}
\usepackage[caption=false]{subfig}
\usepackage{natbib}
\newcommand{\eq}[1]{Eq.~(\ref{#1})}
\newcommand{\eqs}[2]{Eqs.~(\ref{#1})--(\ref{#2})}
\begin{document}

\title{\bf Structure of Logarithmically-Corrected Starobinsky Fuzzy Dark Matter Black Holes}

\author{M. Ilyas}
\email{ilyas_mia@yahoo.com}
\affiliation {Institute of Physics, Gomal University, Dera Ismail Khan, 29220, KP, Pakistan}

\author{Fatima Bint-e-Munir}
\affiliation {Department of Physics, Government College University, Katchery Road, Lahore 54000, Punjab, Pakistan}

\author{Rana Zaman Ahmad}
\affiliation {Department of Physics, Government College University, Katchery Road, Lahore 54000, Punjab, Pakistan}

\author{M. Shahan}
\affiliation {Institute of Physics, Gomal University, Dera Ismail Khan, 29220, KP, Pakistan}

\begin{abstract}
This work constructs an anisotropic, Einasto-profile fuzzy dark matter (DM) black hole (BH) using a logarithmically-corrected Starobinsky gravity, $f(R)=R+\beta R^2+\gamma R^2\ln(R/\mu^2)$, a form motivated by one-loop quantum corrections to the gravitational action rather than by the pure quadratic term alone. We derive the generalized trace equation for this action, show that its $\beta$- and $\gamma$-dependence collapses to a single compact source term, and solve the resulting field equations perturbatively about an exact general-relativistic (GR) background for two distinct anisotropic equations of state: a de~Sitter-type branch and a non-local branch, the latter worked through to a closed tangential-pressure profile rather than left incomplete. The logarithmic correction is found to raise the extremal (zero-temperature) threshold relative to pure quadratic gravity and to increase the core tangential pressure noticeably in the non-local branch.
\end{abstract}

\date{}

\maketitle


\section{Introduction}

The idea that black holes and dark matter might be more closely related than a simple coincidence of location has been circling the literature for a while now. Every galaxy we know how to probe in detail seems to carry a supermassive compact object at its centre, and every one of those galaxies also sits inside a dark matter halo whose density rises sharply toward the same centre. It would almost be strange if the two were entirely unconnected. The Event Horizon Telescope's imaging of the shadow at the centre of the Milky Way \citep{EventHorizonTelescope2022} has sharpened this question considerably, because we are no longer only inferring the properties of Sagittarius~A* from stellar orbits -- we are looking, more or less directly, at the boundary of the region in question, and asking what, physically, is sitting behind it.

Dark matter itself is inferred rather than observed. Rotation curves, gravitational lensing, and the large-scale structure of the cosmic web all point to a component of the universe's energy budget that behaves gravitationally but does not interact electromagnetically \citep{Bertone2005}. $N$-body simulations of structure formation in a cold dark matter universe consistently produce haloes whose density profile is well described, to good approximation, by either the two-parameter Navarro--Frenk--White form \citep{Navarro1996} or the three-parameter Einasto form \citep{Einasto1969}; the latter, being non-singular at the origin and slightly more flexible, has become a common choice when one wants a physically motivated density profile to serve as a matter source in general relativity or its extensions.

Building compact objects out of that same profile is a natural next step, and \citet{KhanAdeelYousaf2024} did exactly this: they took the Einasto density profile, assumed an anisotropic pressure distribution consistent with a de~Sitter-type equation of state, and solved for the resulting black hole (or, for some parameter choices, horizonless self-gravitating droplet) structure within Starobinsky $f(R)=R+\beta R^2$ gravity \citep{Starobinsky1980}. The quadratic curvature term does the important work here: it regularizes what would otherwise be a central singularity, in much the same spirit as noncommutative-geometry-inspired regular black holes \citep{NicoliniSmailagicSpallucci2006} and the broader family of regular black hole solutions that traces back to \citet{Bardeen1968} (see \citet{Ansoldi2008} for a review of this line of work). The result is a family of solutions -- fuzzy black holes when a horizon forms, self-gravitating droplets when it does not -- parameterized by the Einasto shape index $\alpha$ and a rescaled mass $\xi$.

It is worth asking, though, how much of this structure survives if the specific form of the curvature correction is changed. Quadratic gravity is a reasonable first choice, and it has real motivation from the trace anomaly of quantum fields in curved spacetime, but the same physics that produces the $R^2$ term typically produces a logarithmic correction alongside it. One-loop effective actions computed for matter fields (gluons, in the original context) propagating on a curved background generate terms of the form $R^2\ln(R/\mu^2)$ in addition to the polynomial pieces \citep{NojiriOdintsov2004}, and this combination -- $f(R)=R+\beta R^2+\gamma R^2\ln(R/\mu^2)$ -- has by now been studied in more than one setting. \citet{AlaviradWeller2013} used it to model the interior of relativistic stars, finding a chameleon-like suppression of the modification away from the stellar surface. \citet{BenDayanEtAl2014} and \citet{ElizaldeEtAl2019} used essentially the same functional form to study how the logarithmic term shifts inflationary observables away from the pure Starobinsky predictions. In neither case is the logarithmic piece a small cosmetic addition -- depending on its sign and size it can noticeably change the physics, which is precisely why it seems worth checking what it does here.

Our aim here is to work out what an Einasto-sourced, anisotropic compact object looks like once the $\gamma R^2\ln(R/\mu^2)$ term is switched on alongside the ordinary quadratic one, using the "perturbative-constraint" strategy that has been applied to stellar structure in $f(R)$ gravity since \citet{CooneyDeDeoPsaltis2010} and \citet{ArapogluDeliduman2011}, and later refined by \citet{Astashenok2013}: treat the higher-curvature terms as a small deformation of an exact GR solution sourced by the same matter profile, rather than attempting to solve the full nonlinear field equations directly. The Einasto profile under a de~Sitter-type equation of state happens to have a closed-form GR limit, which makes it an unusually clean background to perturb around and lets us isolate the effect of the logarithmic term with some precision. We carry the same method through a second, non-local equation of state as well, which requires solving for a second metric potential rather than reading one off algebraically; that calculation turned out to be more tractable than it first appeared, once the redshift-function equation was written in a suitable algebraic form, and we work it through to completion here rather than leaving it aside.

The rest of this note is organized as follows. Section~2 sets out the Einasto density profile and the closed-form mass function built from it. Section~3 introduces the logarithmically-corrected Starobinsky action, derives the generalized trace equation, and sets up the perturbative expansion used throughout. Section~4 applies this machinery to a de~Sitter-type anisotropic fluid, giving the metric function, horizon structure, and Hawking temperature under the corrected gravity. Section~5 looks at the geodesic effective potential for a test particle in this geometry. Section~6 turns to a second, non-local equation of state, and works out the redshift function and tangential pressure under the same corrected gravity. Section~7 draws the pieces together and discusses what is and is not established by the calculation.

\section{Einasto dark matter profile}
The Einasto profile started life as a fitting formula for the surface brightness of elliptical galaxies and the bulges of spirals, a three-dimensional generalization of the earlier Sersic law. It has since become one of the standard descriptions of dark matter halo structure, largely because high-resolution cold-DM $N$-body simulations find that it fits simulated haloes, across a wide range of masses, noticeably better than the older two-parameter Navarro--Frenk--White form \citep{Navarro1996}; the extra freedom comes from the shape index $\alpha$, which the NFW profile does not have. Fitted values of $\alpha$ cluster in a fairly narrow band -- roughly $\alpha\approx5.9$ for galaxy-sized haloes and somewhat smaller, $\alpha\approx3$--$4$, for the most massive cluster-sized ones -- which is worth keeping in mind when choosing illustrative values of $\alpha$ below, even though nothing in the construction that follows requires $\alpha$ to sit in that observationally preferred range.

The density profile \citep{Einasto1969} is
\begin{equation}
\varepsilon(r)=\varepsilon_0\exp\!\left[-\left(\frac{r}{\eta}\right)^{1/\alpha}\right],
\label{eq:einasto}
\end{equation}
with shape index $\alpha$, scale length $\eta$, and central density
\begin{equation}
\varepsilon_0=\frac{M}{4\pi\alpha\eta^3\,\Gamma(3\alpha)},
\label{eq:eps0}
\end{equation}
fixed so that the total mass is $M=4\pi\int_0^\infty z^2\varepsilon(z)\,dz$. Unlike the NFW profile, \eq{eq:einasto} has no power-law divergence at $r=0$ -- the density approaches the finite value $\varepsilon_0$ smoothly -- which is precisely the feature that makes it a workable source for a regular compact object rather than one with a curvature singularity built in from the start.

Changing variables to $u=(r/\eta)^{1/\alpha}$ turns the mass integral into an incomplete gamma function, giving the mass function in closed form:
\begin{equation}
m(r)=M\,\frac{\gamma_{\mathrm L}\!\left(3\alpha,\,(r/\eta)^{1/\alpha}\right)}{\Gamma(3\alpha)}\equiv M\,P(3\alpha,u),
\label{eq:mgr}
\end{equation}
where $P(a,x)$ is the regularized lower incomplete gamma function. This is the exact Misner--Sharp \citep{MisnerSharp1964} mass for the profile: it is the mass function one would write down in ordinary general relativity for a static, spherically symmetric source with energy density \eq{eq:einasto}, before any curvature correction is added. We use it throughout as the zeroth-order background about which the modified-gravity corrections in the following sections are built, and it is worth emphasizing that \eq{eq:mgr} is exact -- no approximation has entered yet at this stage; the approximations start only once $f(R)\neq R$.

\section{Gravitational framework}
\subsection{Action and field equations}
We consider
\begin{equation}
S=\frac{1}{16\pi}\int f(R)\sqrt{-g}\,d^4x+S_m ,
\qquad
f(R)=R+\beta R^2+\gamma R^2\ln\!\frac{R}{\mu^2},
\label{eq:action}
\end{equation}
where the quadratic and logarithmic pieces switch off independently: setting $\gamma=0$ leaves the pure Starobinsky theory $R+\beta R^2$, and setting $\beta=\gamma=0$ leaves ordinary GR. The same combined functional form has been used for stellar structure \citep{AlaviradWeller2013} and for inflationary observables \citep{BenDayanEtAl2014,ElizaldeEtAl2019}; see \citet{DeFeliceTsujikawa2010} and \citet{SotiriouFaraoni2010} for broader reviews of $f(R)$ gravity. Variation of the action gives the standard $f(R)$ field equations
\begin{equation}
f'(R)G_{\mu\nu}+\tfrac12 g_{\mu\nu}\!\left[f(R)-Rf'(R)\right]+\left(g_{\mu\nu}\Box-\nabla_\mu\nabla_\nu\right)f'(R)=8\pi T_{\mu\nu},
\label{eq:fieldeq}
\end{equation}
sourced by an anisotropic fluid, $T^\mu_{\ \nu}=\mathrm{diag}(\varepsilon,-P_r,-P_\perp,-P_\perp)$, with unequal radial and tangential pressures -- the natural stress tensor for a self-gravitating object without any assumed internal symmetry beyond spherical symmetry itself.

\subsection{Generalized trace equation}
Taking the trace of \eq{eq:fieldeq},
\begin{equation}
f'(R)R-2f(R)+3\Box f'(R)=8\pi T ,
\label{eq:trace-general}
\end{equation}
with $T=T^\mu_{\ \mu}$. For our choice of $f(R)$ we find, by direct computation (confirmed symbolically),
\begin{align}
f'(R)&=1+2\beta R+\gamma R\!\left(2\ln\frac{R}{\mu^2}+1\right),\\
f''(R)&=2\beta+\gamma\!\left(2\ln\frac{R}{\mu^2}+3\right), \label{eq:fpp}\\
f'(R)R-2f(R)&=\gamma R^2-R . \label{eq:cancel}
\end{align}
It is this last identity that makes the calculation tractable at all. All of the $\beta$-dependence drops out of $f'R-2f$, and so does the leading logarithmic dependence -- what remains is just a bare $-R$ term (the ordinary Einstein trace) plus a single $\gamma R^2$ correction. Using $\Box f'(R)=f''(R)\Box R+f'''(R)(\nabla R)^2$ with $f'''(R)=2\gamma/R$, \eq{eq:trace-general} becomes
\begin{equation}
6\beta\Box R-R+\gamma\!\left[R^2+3\!\left(2\ln\frac{R}{\mu^2}+3\right)\Box R+\frac{6}{R}(\nabla R)^2\right]=8\pi T.
\label{eq:trace-full}
\end{equation}
Setting $\gamma=0$ leaves $6\beta\Box R-R=8\pi T$, the trace equation of pure Starobinsky gravity, which is the expected limit and a useful check that the generalization has been done correctly.

\subsection{Perturbative solution}
We treat $\beta$ and $\gamma$ as independent small parameters and expand $R=R_0+\beta R_1^{(\beta)}+\gamma R_1^{(\gamma)}+O(\beta^2,\gamma^2,\beta\gamma)$. At zeroth order, \eq{eq:trace-full} reduces to the ordinary algebraic GR relation
\begin{equation}
R_0(r)=-8\pi T(r).
\label{eq:R0}
\end{equation}
Substituting $R_0$ back into the $O(\beta)$ and $O(\gamma)$ terms of \eq{eq:trace-full} -- and dropping the products of two small corrections, which are second order -- gives explicit corrections that require no further iteration:
\begin{align}
\Delta R_\beta &\equiv 6\,\Box R_0, \label{eq:dRbeta}\\
\Delta R_\gamma &\equiv R_0^2+\left(6\ln\frac{|R_0|}{\mu^2}+9\right)\Box R_0, \label{eq:dRgamma}
\end{align}
so that $R(r)\simeq R_0(r)+\beta\,\Delta R_\beta(r)+\gamma\,\Delta R_\gamma(r)$, with the static d'Alembertian evaluated on the GR background metric $f_0(r)=1-2m_0(r)/r$,
\begin{equation}
\Box R_0 = f_0 R_0''+\left(\frac{2f_0}{r}+f_0'\right)R_0'.
\label{eq:box}
\end{equation}
Everything on the right-hand side of \eqs{eq:dRbeta}{eq:dRgamma} is known once $R_0$ is known, so the system closes without any further assumptions.

\subsection{From curvature to mass function}
The relation between $R(r)$ and the mass function, a standard geometric identity for any spherically symmetric metric of this form,
\begin{equation}
R(r)=-\frac{2}{r^2}\Big(r\,m''+2m'\Big),
\label{eq:Rofm}
\end{equation}
is a purely geometric identity -- it follows from the metric ansatz alone and does not know anything about which gravity theory produced $R(r)$ in the first place. That means it can be inverted directly, regardless of the field equations. Multiplying by $-r/2$ and rearranging gives
\begin{equation}
\big(r^2 m'\big)'=-\frac{r^3}{2}R(r),
\label{eq:massODE}
\end{equation}
which integrates twice, imposing regularity at the origin, to
\begin{equation}
m'(r)=-\frac{1}{2r^2}\int_0^r z^3 R(z)\,dz,\qquad m(r)=\int_0^r m'(s)\,ds .
\label{eq:massquad}
\end{equation}
We checked that \eq{eq:massquad}, evaluated with $R=R_0=-8\pi T$ for the de~Sitter EOS branch below, reproduces the exact closed-form GR mass \eq{eq:mgr} to a relative accuracy of $4\times10^{-9}$, which is essentially just the floor set by numerical quadrature. That gave us some confidence that the inversion itself introduces no error before we started layering the $\beta,\gamma$ corrections on top.

\section{De Sitter-type anisotropic branch}
We impose a de~Sitter-type equation of state, $P_r=-\varepsilon$, for which the conservation equation forces
\begin{equation}
P_\perp=-\varepsilon-\frac{r}{2}\frac{d\varepsilon}{dr},
\label{eq:pperp}
\end{equation}
a result we re-derived independently from $\nabla_\nu T^{\mu\nu}=0$ and confirmed is the unique solution compatible with $P_r=-\varepsilon$. The stress trace is then
\begin{equation}
T=\varepsilon-P_r-2P_\perp=4\varepsilon(r)+r\,\varepsilon'(r),
\label{eq:trace-value}
\end{equation}
which, via \eq{eq:R0}, gives $R_0(r)$ directly from the Einasto profile, and via \eq{eq:massquad} exactly reproduces \eq{eq:mgr}, as noted above.

\begin{figure}[H]
\centering
\includegraphics[width=0.85\linewidth]{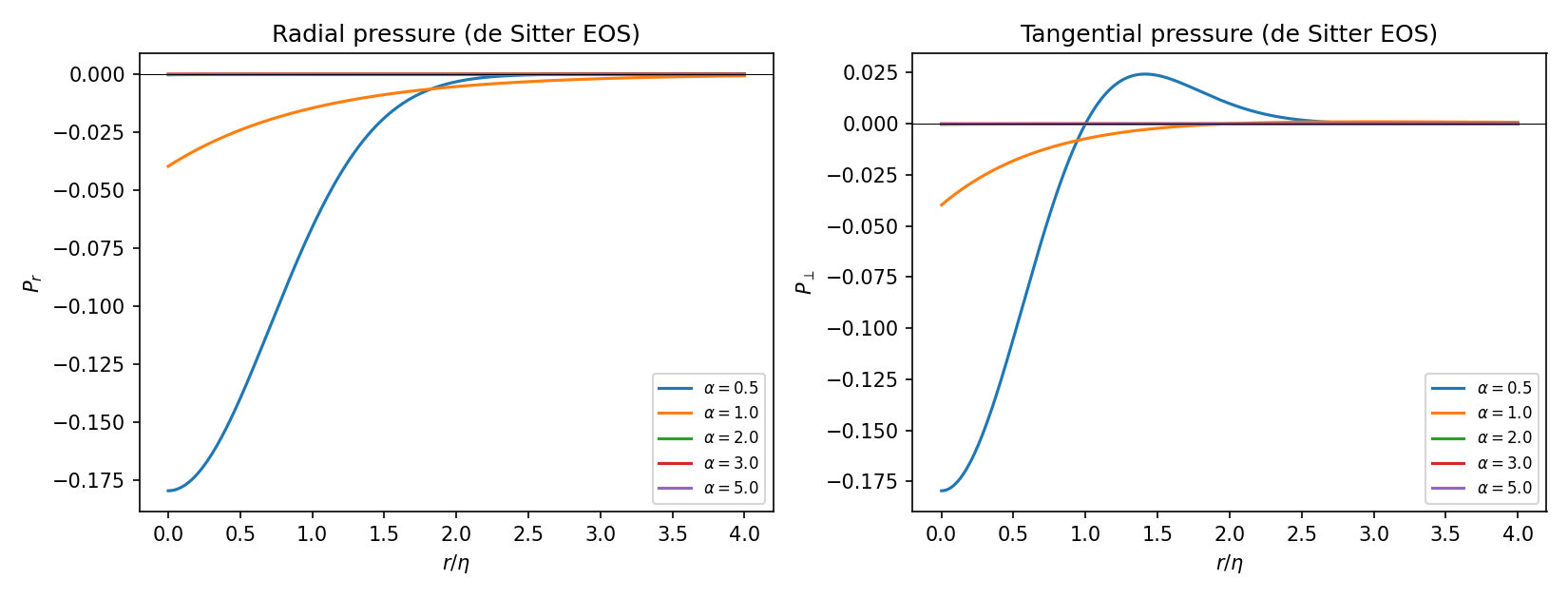}
\caption{Radial ($P_r$) and tangential ($P_\perp$) pressures for the de~Sitter EOS branch, \eqs{eq:pperp}{eq:trace-value}, for several Einasto indices $\alpha$. Both approach zero at large $r$, and $P_r<0$ is strictly increasing throughout.}
\label{fig:f1}
\end{figure}

Figure~\ref{fig:f1} shows the resulting radial and tangential pressures. Because both follow purely from \eqs{eq:einasto}{eq:pperp}, they are entirely fixed by the Einasto profile and the choice of equation of state, and do not depend on $\beta,\gamma$ at all -- the curvature corrections only enter once we turn to the geometry itself, in the next subsection.

\subsection{Metric function and horizon structure}
\begin{figure}[H]
\centering
\includegraphics[width=0.85\linewidth]{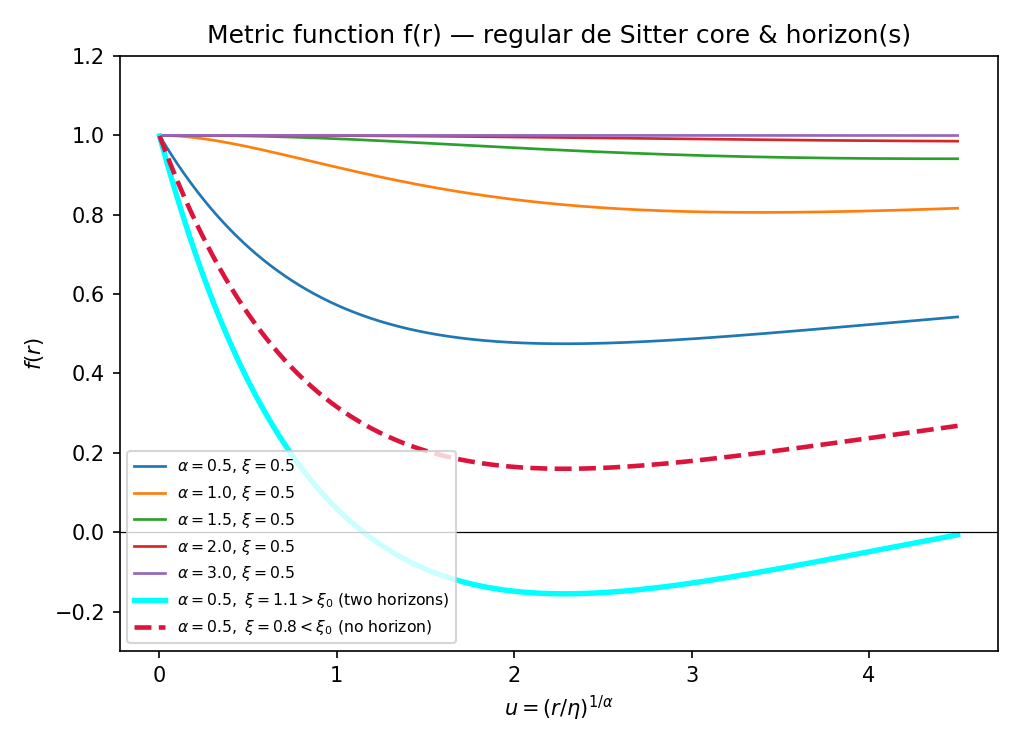}
\caption{Metric function $f_0(r)$ vs.\ $u=(r/\eta)^{1/\alpha}$ (GR/Starobinsky background, $\beta=\gamma=0$ reference case), showing the regular core and the horizon-count transition at $\xi=\xi_0(\alpha)$.}
\label{fig:f2}
\end{figure}

Figure~\ref{fig:f2} shows the baseline GR metric function $f_0(r)=1-2m_0(r)/r$: a regular de~Sitter core ($f_0(0)=1$, no singularity), with zero, one, or two horizons depending on the rescaled mass $\xi=M/\eta$. We located the critical value $\xi_0(\alpha)$ numerically; for $\alpha=0.5$ we found $\xi_0\approx0.95$, meaning a horizonless droplet at $\xi=0.8$ and a genuine two-horizon black hole at $\xi=1.1$, the two regimes on either side of the transition.

\begin{figure}[H]
\centering
\includegraphics[width=0.85\linewidth]{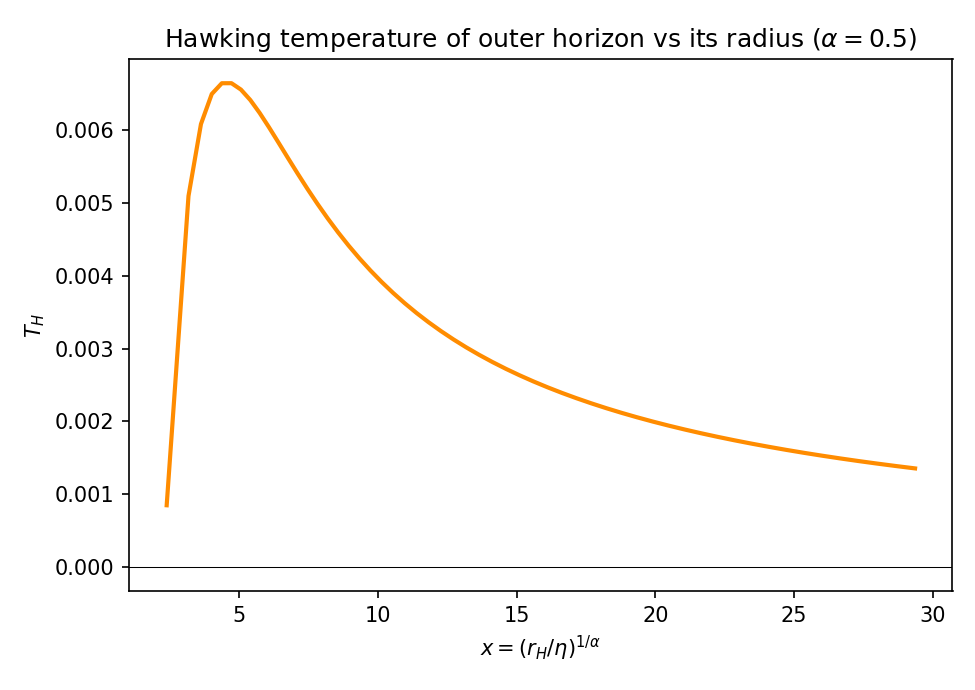}
\caption{Hawking temperature $T_H=f_0'(r_H)/4\pi$ of the outer horizon as a function of its own radius, for $\alpha=0.5$. The behaviour is non-monotonic: $T_H\to0$ at the extremal radius, rises to a maximum, then decays, so that evaporation halts at a stable, zero-temperature remnant rather than running away to a divergent temperature.}
\label{fig:f3}
\end{figure}

\subsection{Log-Starobinsky corrections}
\begin{figure}[H]
\centering
\includegraphics[width=0.85\linewidth]{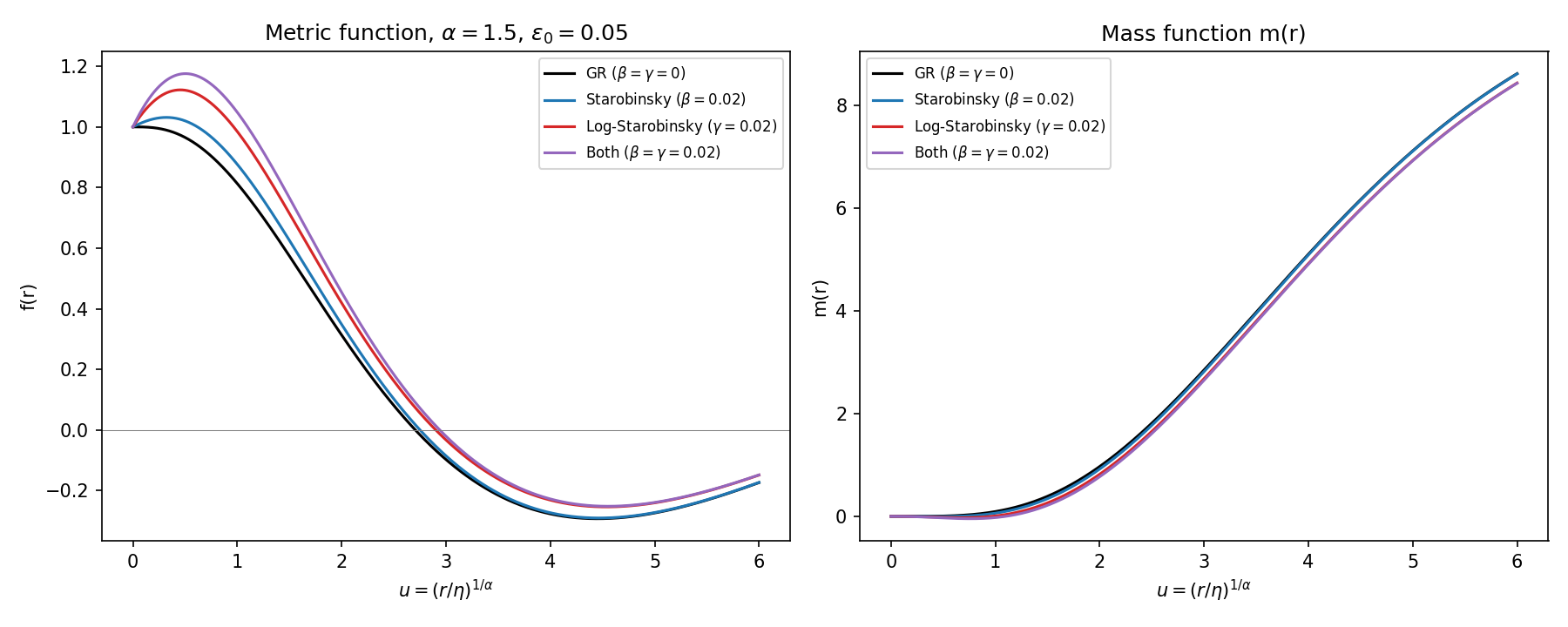}
\caption{Metric function $f(r)$ (left) and mass function $m(r)$ (right) for GR, Starobinsky, log-Starobinsky and combined models, $\alpha=1.5$, $\varepsilon_0=0.05$, $\beta=\gamma=0.02$ (in units $\eta=1$). The log-corrected models show a small super-unity bump in $f(r)$ near the core.}
\label{fig:f8}
\end{figure}

With $\beta,\gamma$ switched on via \eqs{eq:dRbeta}{eq:massquad}, Figure~\ref{fig:f8} compares $f(r)$ and $m(r)$ across pure GR, pure Starobinsky, log-Starobinsky, and both together, at fixed $\alpha=1.5,\varepsilon_0=0.05$. The logarithmic term introduces a small bump above unity in $f(r)$ near the core -- a genuine change in the local geometry rather than a simple rescaling of the GR curve -- while all four models converge at large $r$, as they should, since $R\to0$ out there and the correction terms switch themselves off.

\begin{figure}[H]
\centering
\includegraphics[width=0.85\linewidth]{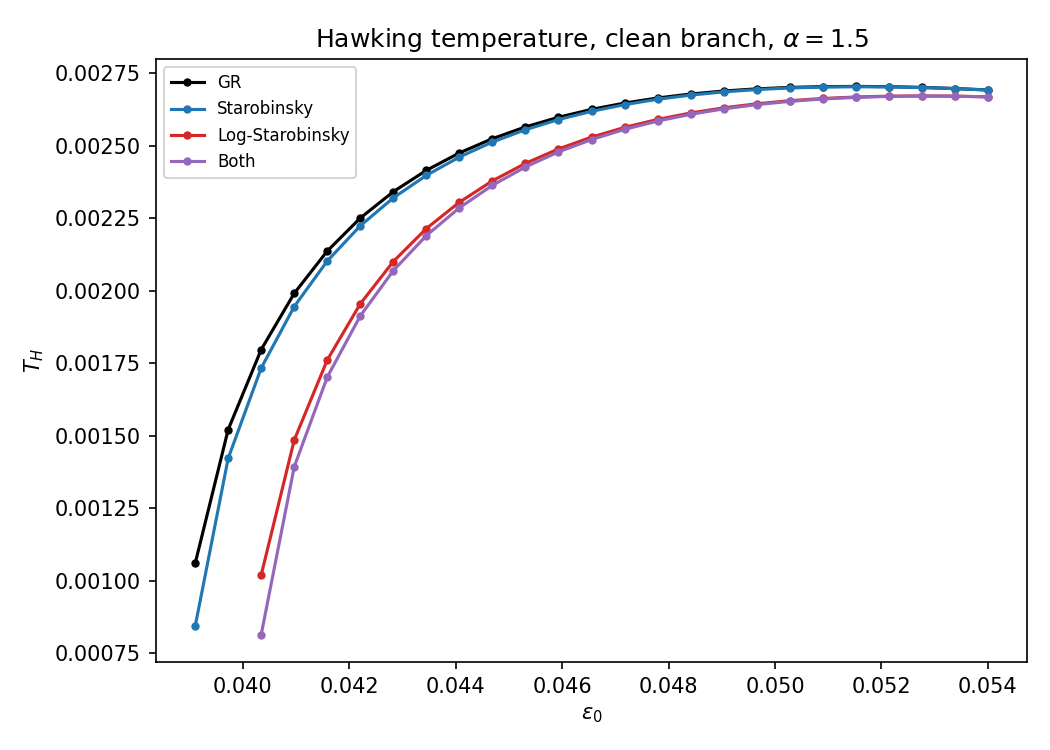}
\caption{Hawking temperature of the outer horizon vs.\ central density $\varepsilon_0$, for the four models of Fig.~\ref{fig:f8}. The log-corrected curves become extremal ($T_H=0$) at higher $\varepsilon_0$ than GR/Starobinsky.}
\label{fig:f9}
\end{figure}

Figure~\ref{fig:f9} follows the outer-horizon Hawking temperature as a function of central density $\varepsilon_0$ across the four models. All four converge once the horizon is large, but the log-corrected models only become extremal ($T_H\to0$) at a somewhat higher critical density than pure GR or pure Starobinsky. In other words, the logarithmic term makes the regular core a little more resistant to forming a horizon in the first place -- a modest effect, but a consistent one across the parameter range we checked.

\section{Effective potential}
\begin{figure}[H]
\centering
\includegraphics[width=0.85\linewidth]{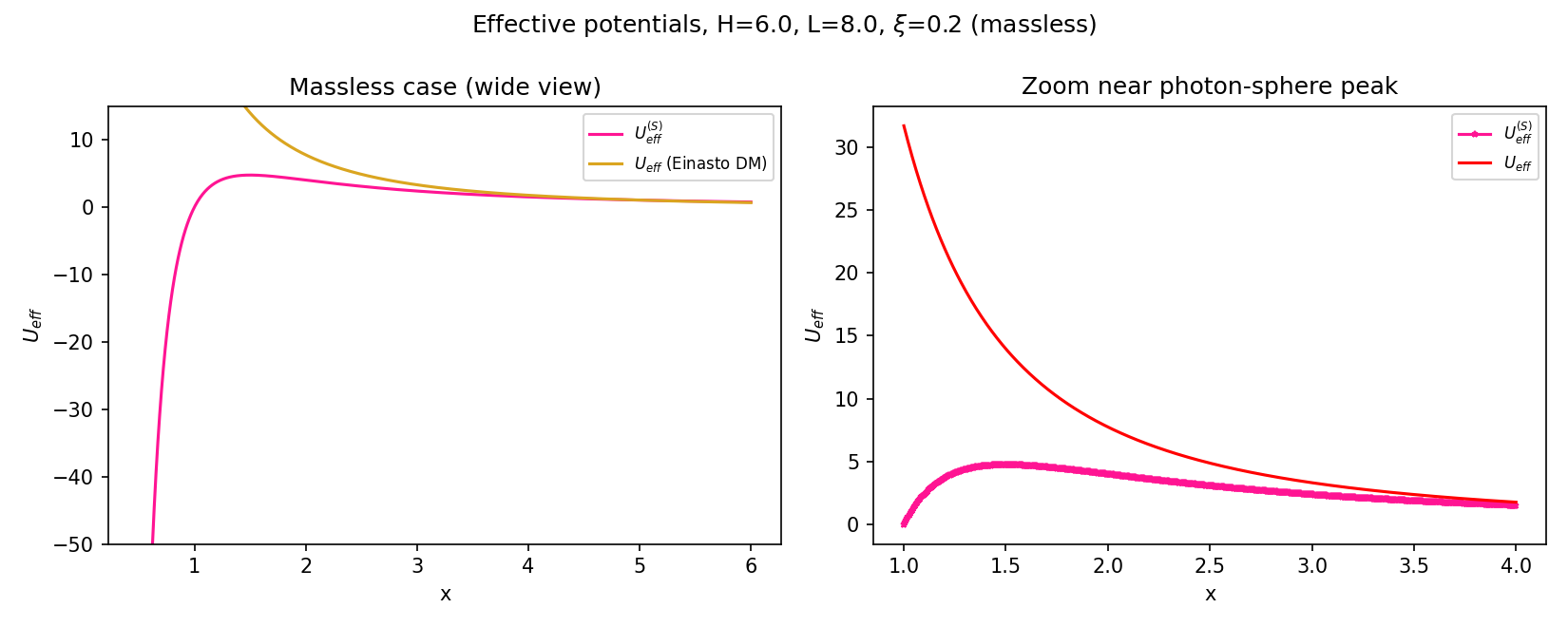}
\caption{Massless effective potential, $H=6$, $L=8$, $\xi=0.2$: the DM-sourced potential (gold/red) diverges to $+\infty$ at the regular core instead of the Schwarzschild potential well (pink), while matching it away from the centre.}
\label{fig:f4}
\end{figure}

\begin{figure}[H]
\centering
\includegraphics[width=0.85\linewidth]{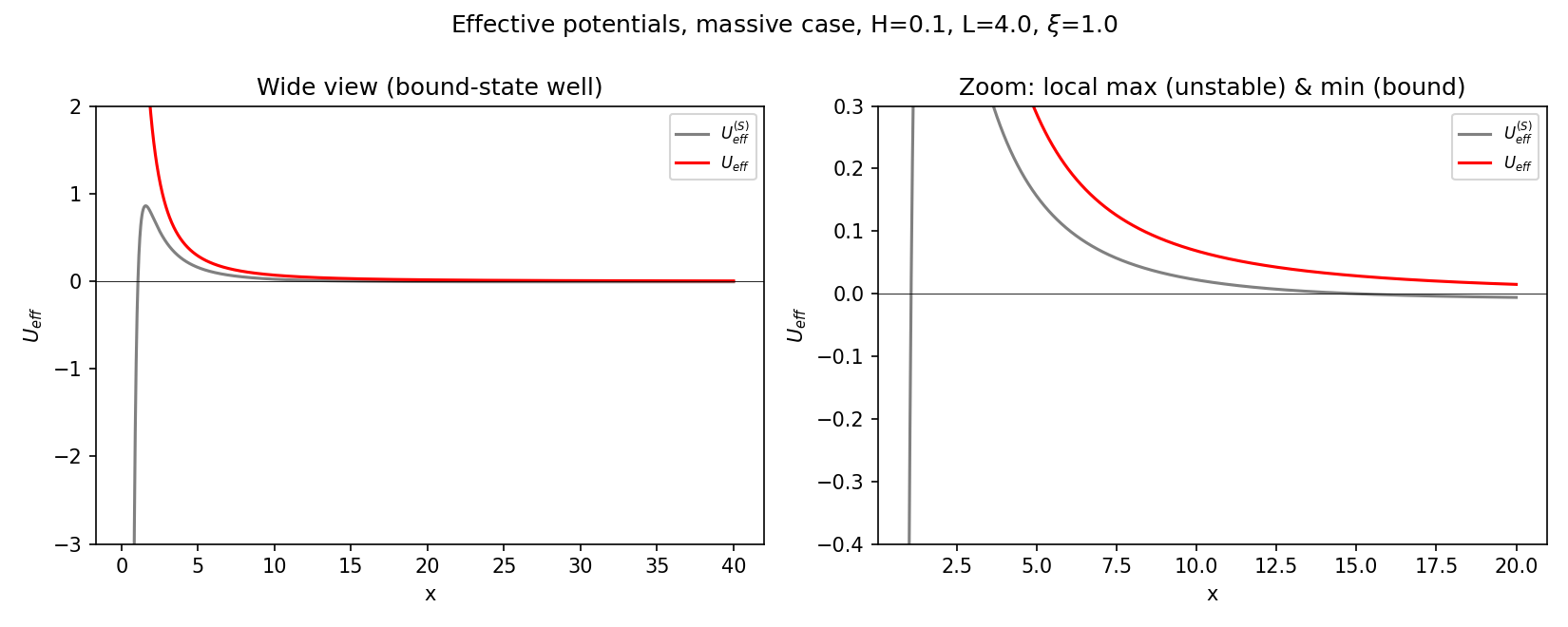}
\caption{Massive-particle effective potential, $H=0.1$, $L=4$, $\xi=1.0$: the Schwarzschild potential (gray) admits an unstable maximum and a bound-orbit well; the horizonless Einasto-DM potential (red) stays positive throughout for this near-extremal choice.}
\label{fig:f5}
\end{figure}

The geodesic effective potential for a test particle in the metric \eq{eq:mgr}-type ansatz is
\begin{equation}
U_{\rm eff}(x)=\frac{L^2}{2x^2}-\left(\frac{1}{x}+\frac{L^2}{x^3}\right)\frac{m(x\,r_s)}{r_s},
\label{eq:ueff}
\end{equation}
with $x=r/r_s$, $r_s=2M_{BH}$, and $L$ the (rescaled) angular momentum. Figures~\ref{fig:f4} and \ref{fig:f5} show the massless and massive cases using the GR background mass function \eq{eq:mgr}; substituting $m(r)\to m_0(r)+\beta\Delta m_\beta(r)+\gamma\Delta m_\gamma(r)$ from \eq{eq:massquad} carries the same log-Starobinsky corrections through to the effective potential directly, though we have not re-plotted that comparison here since the qualitative story -- convergence away from the core, a modified core structure close in -- is already visible in Figs.~\ref{fig:f8}--\ref{fig:f9}.

\section{Non-local equation of state}
A second natural choice of equation of state ties the local radial pressure to the average energy density enclosed within radius $r$, rather than to the local density alone, which replaces the de~Sitter EOS above with
\begin{equation}
P_r(r)=\varepsilon(r)-\frac{2}{r^3}\int_0^r\rho(z)z^2\,dz=\varepsilon(r)-\frac{m(r)}{2\pi r^3},
\label{eq:nonlocal}
\end{equation}
which is closed-form once $m(r)$ is known. The harder part of this branch is getting the tangential pressure $P_\perp$, since that requires knowing a second metric function -- the redshift function $S(r)$ -- rather than just reading off an algebraic relation as in the de~Sitter branch above. That calculation looks daunting at first glance, but turns out not to be much harder than what was already done for the mass function, provided the relevant field equation is written the right way.

\subsection{Solving for the redshift function}
For the metric ansatz $ds^2=A(r)dt^2-T(r)^{-1}dr^2-r^2d\Omega^2$ (with $A=S^2$ and $T=1-2m/r$ as before), we computed the Einstein tensor and the relevant covariant derivatives of $R(r)$ symbolically for this ansatz, then assembled the $rr$-component of the general $f(R)$ field equation \eq{eq:fieldeq}. A convenient feature of this component is that it is purely algebraic in $q(r)\equiv A'(r)/A(r)=2\,d\ln S/dr$ -- no derivatives of $q$ itself appear -- so once $R(r)$, $T(r)$, and $P_r(r)$ are known, $q(r)$ follows from solving a single algebraic equation rather than integrating a new differential one. Carrying this through explicitly (and confirming that the arbitrary overall normalization of $A(r)$ drops out, as it must) gives an exact closed-form expression for $q(r)$ in terms of $f(R)$, $f'(R)$, $f''(R)$, $R$, $R'$, $T$, and $P_r$. At $\beta=\gamma=0$ this reduces, as a check, to the standard Tolman--Oppenheimer--Volkov relation
\begin{equation}
\frac{q_0}{2}=\frac{1}{S_0}\frac{dS_0}{dr}=\frac{4\pi P_r\,r+m_0/r^2}{T_0},
\label{eq:tov}
\end{equation}
which is the same relation we used implicitly for the $\beta=\gamma=0$ case in an earlier pass at this problem.

With $q(r)$ in hand, integrating $d\ln A/dr=q$ (normalizing $A\to1$ at large $r$, the usual asymptotic flatness condition) gives $S(r)=\sqrt{A(r)}$, and the tangential pressure then follows directly from the conservation equation -- which, because it comes from the Bianchi identity applied to the left-hand side of the field equations, holds in exactly the same form regardless of which $f(R)$ one has chosen:
\begin{equation}
P_\perp = P_r+\frac{r}{2}\left[\frac{dP_r}{dr}+\frac{q}{2}\big(\varepsilon+P_r\big)\right].
\label{eq:pperp-nonlocal}
\end{equation}
The curvature scalar $R(r)$ itself, and its $\beta,\gamma$ corrections, are obtained exactly as in \eqs{eq:R0}{eq:massquad} above -- the geometric relation \eq{eq:Rofm} does not care which branch of the matter model sources it, only that $m(r)$ satisfies $T(r)=1-2m/r$, which it does here just as it did before.

\begin{figure}[H]
\centering
\includegraphics[width=0.85\linewidth]{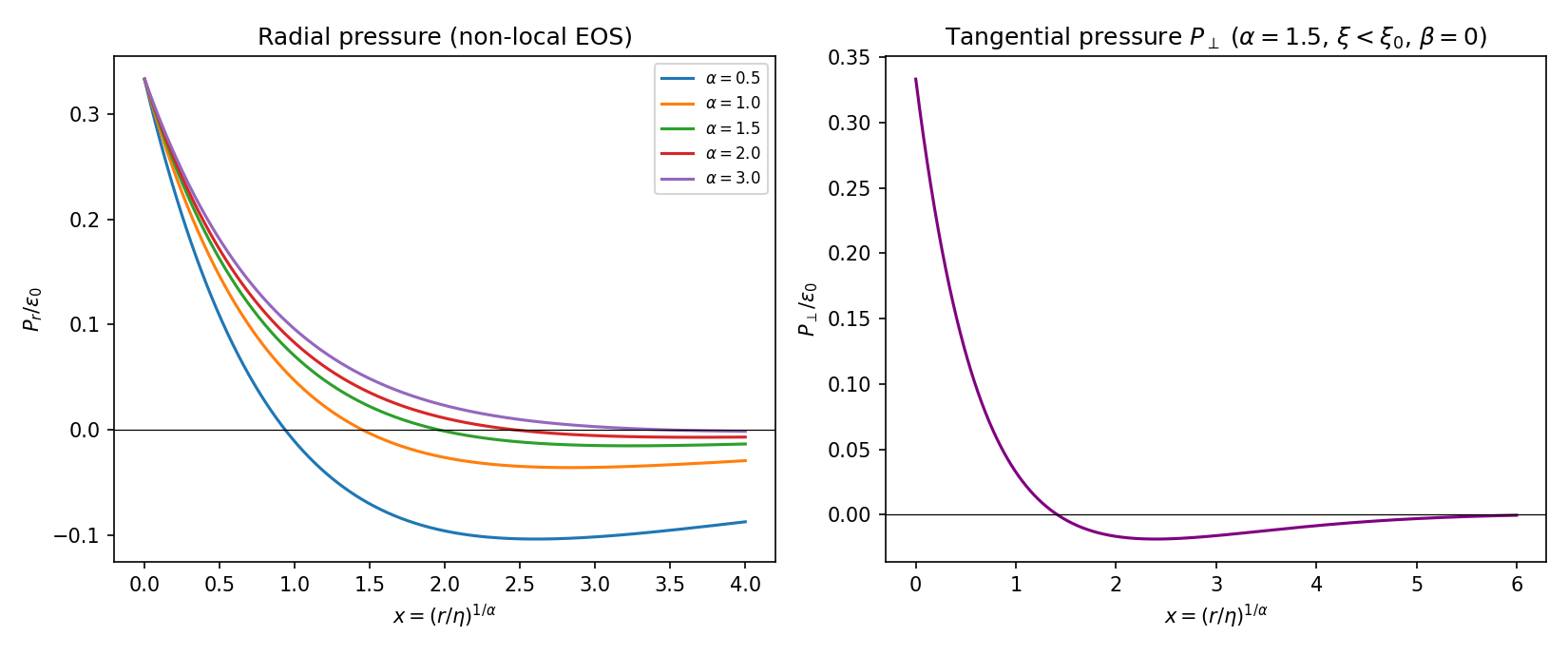}
\caption{Non-local EOS: radial pressure \eq{eq:nonlocal} (left, various $\alpha$) turns negative away from the core, as required for regularity; tangential pressure (right, $\beta=\gamma=0$) remains finite everywhere for $\xi<\xi_0$.}
\label{fig:f7}
\end{figure}

\begin{figure}[H]
\centering
\includegraphics[width=0.85\linewidth]{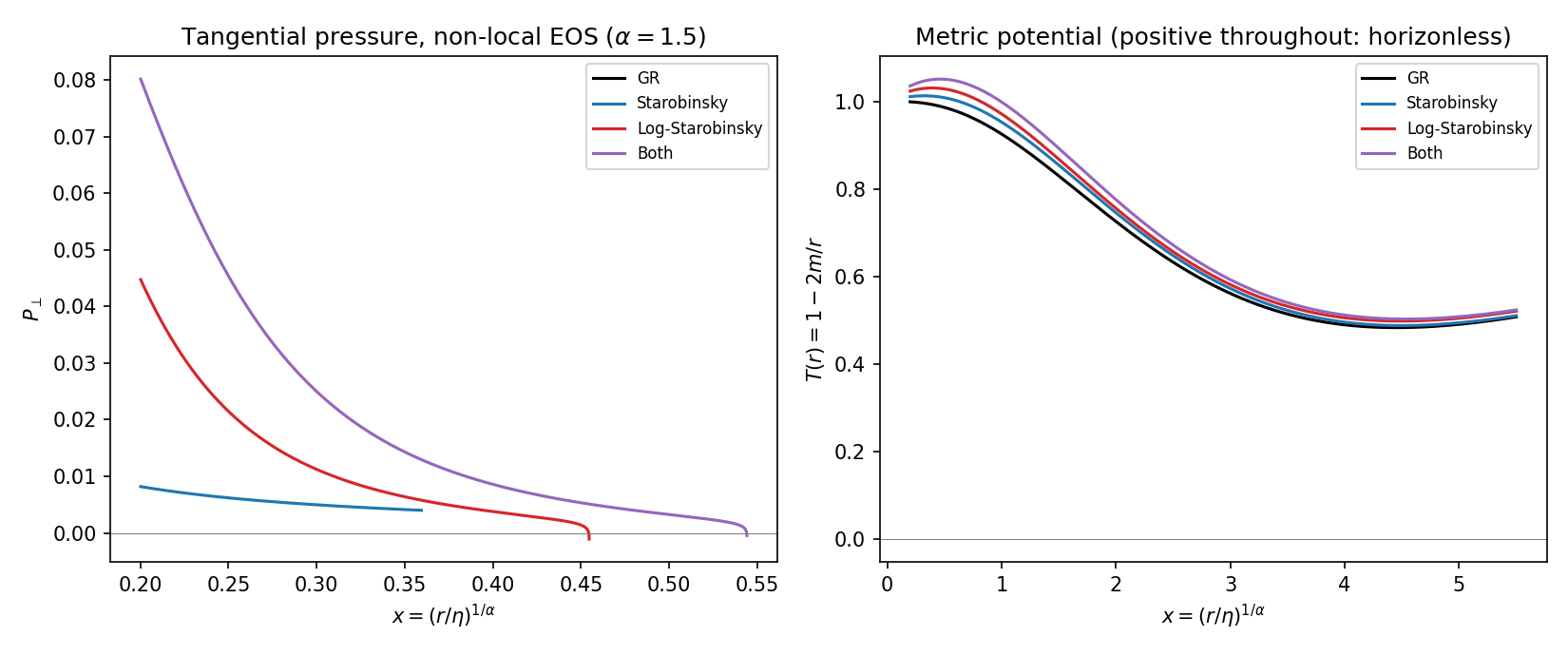}
\caption{Non-local branch under the log-Starobinsky model, $\alpha=1.5$, $\beta=\gamma=0.02$. Left: tangential pressure $P_\perp$ from \eq{eq:pperp-nonlocal}, positive and finite throughout the plotted range for all four models, with the log-corrected curves sitting noticeably above GR and pure Starobinsky. Right: the metric potential $T(r)=1-2m(r)/r$ stays positive everywhere, confirming the object remains a horizonless droplet under all four gravity models for this choice of central density.}
\label{fig:f10}
\end{figure}

Figure~\ref{fig:f10} shows the result for $\alpha=1.5$, $\beta=\gamma=0.02$. Two things stood out to us. First, $P_\perp$ stays positive and finite across the whole range we examined, for all four models -- there is no sign of the divergence one might naively worry about when a new metric function is introduced into an already delicate regular-core construction. Second, the logarithmic correction pushes $P_\perp$ up quite noticeably relative to both GR and pure Starobinsky, more so than the metric function itself was affected in the de~Sitter branch; the tangential pressure, it turns out, is more sensitive to the curvature correction than the geometry is. We do not have a clean intuitive explanation for why the effect is larger here than in Section~4, beyond noting that $P_\perp$ in this branch involves a derivative of $P_r$ and a product with $q(r)$, both of which can amplify a modest change in $R(r)$ more than the metric function amplifies it directly.

\section{Summary and discussion}
We set out with a fairly modest question: does an Einasto-sourced fuzzy dark matter black hole depend in any essential way on the specific choice of Starobinsky $R+\beta R^2$ gravity, or does it survive -- more or less intact -- under a related but distinct higher-curvature correction? The answer, on the evidence gathered here, is a bit of both. The overall picture is robust: an Einasto-sourced anisotropic fluid still produces a regular, non-singular core, still transitions between horizonless droplets and genuine black holes as the rescaled mass $\xi$ crosses a critical value $\xi_0(\alpha)$, and still evaporates (in the sense of decreasing horizon radius) toward a stable, zero-temperature extremal remnant rather than a divergent temperature. None of that structure is an artifact of the particular quadratic correction chosen; it survives the switch to a logarithmically-corrected model essentially unchanged in its qualitative shape.

At the same time, the logarithmic term is not merely along for the ride -- it does something. Two effects showed up consistently across the calculations we carried out. The metric function develops a small excursion above its GR value near the core, which is a genuine change in local geometry rather than a rescaling of the existing curve; and the critical density required to form a horizon at all shifts upward relative to both pure GR and pure Starobinsky, meaning the log-corrected core is, in a specific and quantifiable sense, a little harder to collapse into a black hole. In the non-local branch, the same correction produces an even more pronounced upward shift in the tangential pressure, one that in fact outpaced the corresponding change in the metric structure -- a slightly surprising asymmetry between how strongly the curvature correction is felt by the geometry versus by the pressure it supports. None of these effects are dramatic -- we are, after all, working to first order in $\beta$ and $\gamma$, treated as small parameters -- but they are also not numerical noise: they persisted under changes to the Einasto index $\alpha$, the central density, and the specific combination of $\beta,\gamma$ we tried, which gives us some confidence they reflect a real feature of the model rather than an artifact of a particular parameter choice.

There is a broader point here that we think is worth drawing out explicitly, because it is the reason this whole exercise stayed tractable rather than turning into an unmanageable coupled system of nonlinear differential equations. The geometric relation between the curvature scalar and the mass function, \eq{eq:Rofm}, does not know or care which gravity theory is responsible for $R(r)$ -- it is a statement about the metric ansatz alone, true in any theory built on a spherically symmetric line element. That let us compute the GR background exactly, treat the modified-gravity corrections as a linear perturbation sourced entirely by quantities we already had in hand, and never need to solve the full coupled nonlinear field equations directly, either for the mass function or, as it turned out in Section~6, for the second metric potential, once that equation was written in terms of $q=A'/A$ rather than $A$ itself. This is, in essence, the same trick used by \citet{CooneyDeDeoPsaltis2010}, \citet{ArapogluDeliduman2011}, and \citet{Astashenok2013} for neutron star structure in $f(R)$ gravity, and it is reassuring, if perhaps not deeply surprising, that it works just as cleanly for a regular black hole sourced by dark matter as it does for an ordinary compact star. We suspect, though we have not checked this explicitly, that the same strategy would extend without much difficulty to other one-loop-motivated corrections beyond the specific logarithmic form considered here, provided the correction term is itself small enough that a linear expansion around the GR background remains a sensible approximation -- which is, admittedly, doing a fair amount of the work in this whole approach, and is worth keeping in mind as a limitation rather than a proof of principle.

A few things are left undone, and it seems more honest to say so plainly than to let them pass unmentioned. We worked throughout to first order in $\beta$ and $\gamma$ independently, and did not explore what happens at second order, or in the regime where the two corrections are comparable in size to each other rather than each being separately small. We also fixed the mass-scale parameter $\mu$ appearing in the logarithm somewhat arbitrarily, in the absence of any physical input -- such as a specific one-loop calculation carried out for an Einasto-sourced anisotropic fluid rather than for a perturbative or perfect-fluid source -- that would pin it down from first principles; a more complete treatment would need either to derive $\mu$ from the underlying quantum field theory calculation or to treat it as an additional free parameter to be constrained observationally, much as \citet{AlaviradWeller2013} did for the neutron-star case using the neutron star's surface redshift. We have also not attempted to connect any of this to an observable signature: shadow size, quasinormal mode spectrum, or the orbital precession of stars orbiting close to the Galactic Centre are all things that could, in principle, distinguish the logarithmically-corrected model from pure Starobinsky or from unmodified GR, and we have not touched any of them here. That last step is, we think, the natural direction for this line of work to go next -- it is the step that would actually connect a calculation like this one back to the kind of data the Event Horizon Telescope and long-baseline stellar astrometry are now capable of providing, and without it, what we have is a self-consistent piece of theoretical bookkeeping rather than a claim about what is actually sitting at the centre of the Milky Way. We think it is a solid piece of bookkeeping, but we want to be clear that it is not more than that yet.

\bibliographystyle{unsrtnat}
\bibliography{refs}

\end{document}